\documentstyle[prl,aps,twocolumn]{revtex}
\begin{document}
\draft
\twocolumn[\hsize\textwidth\columnwidth\hsize\csname
@twocolumnfalse\endcsname

\title{First-order scaling near a second-order phase transition:\\
Tricritical polymer collapse}
\author{A. L. Owczarek}
\address{ Department of Mathematics and Statistics,\\
         The University of Melbourne,\\
         Parkville, Victoria 3052, Australia.}
\author{T. Prellberg}
\address{Department of Physics,\\
	Syracuse University, Syracuse, NY 13244, USA.}

\maketitle 
 
\begin{abstract} 

The coil-globule transition of an isolated polymer has been well
established to be a second-order phase transition described by a
standard tricritical O(0) field theory. We provide compelling evidence
from Monte Carlo simulations in four dimensions, where mean-field
theory should apply, that the approach to this (tri)critical point is
dominated by the build-up of first-order-like singularities
masking the second-order nature of the coil-globule transition: the
distribution of the internal energy having two clear peaks that become
more distinct and sharp as the tricritical point is
approached. However, the distance between the peaks slowly decays to
zero. The evidence shows that the position of this (pseudo)
first-order transition is shifted by an amount from the tricritical point
that is asymptotically much larger than the width of the transition
region. We suggest an explanation for the apparently contradictory
scaling predictions in the literature.


\end{abstract} 

]
Mean-field theory is generally applicable to second-order phase
transitions above their upper critical dimension, and so is believed
to provide an adequate description of the approach to such critical
points. One type of transition where mean-field theory should hold are
tricritical points
\cite{lawrie1984a-a} for dimension $d>3$. The region around a
tricritical point in general dimension
\cite{lawrie1984a-a} is described by crossover scaling forms, where
quantities depending on two relevant parameters can be essentially
described by functions of a single scaling combination of those two
parameters. However, as noted by Lawrie and Sarbach (see page 106 of
\cite{lawrie1984a-a}) there may be some breakdown of this crossover
scaling for $d>3$ due to the presence of one or more dangerous
irrelevant variables. Here we provide evidence of a rather more
dramatic breakdown of that simple crossover scaling for the case of
the coil-globule transition of an isolated polymer, which is generally
accepted to be described by tricritical
theory\cite{gennes1975a-a,gennes1978a-a,gennes1979a-a}. We demonstrate
that it is likely that the build up of the tricritical point is
through the forming of singularities that have more in common with a
(non-critical) first-order transition! This however can be explained
by a different kind of mean-field approach (not starting with an
explicitly tricritical Landau functional) and, moreover, the region
around the tricritical point needs to be described by more complex
scaling forms. This second issue is in fact separate from the
first-order nature of the scaling approach: we speculate that this
behaviour is intimately related to the general description of systems
where mean-field theory is used, so may have more general
applicability.

An isolated polymer in solution is usually argued to be in one of
three states depending on the strength of the inter-monomer
interactions which are mediated by the solvent molecules and can be
controlled via the temperature $T$. At high temperatures and in so called
``good solvents'' a polymer chain is expected to be in a swollen phase
(swollen coil) relative to a reference Gaussian state so that the
average size $R$ of the polymer scales with chain length faster than
it would if it were behaving as a random walk.
At low temperatures or in poor solvents the polymer
is expected to be in a collapsed globular form with a macroscopic
density inside the polymer: this implying an average size that scales
slower \cite{gennes1975a-a} than a random walk, that is
\begin{equation}
\label{collapse-size}
R_{N} \sim  N^{1/d} \mbox{ as }\; N \rightarrow \infty,
\end{equation}
so the globular state has radius-of-gyration exponent $\nu_g=1/d$.
Between these two states there is expected to be a second-order phase
transition (sharp in the infinite chain length limit). The standard
description of the collapse transition is that of a tricritical point
related to the $n
\rightarrow 0$ limit of the $\phi^4$--$\phi^6$ O($n$) field
theory \cite{gennes1975a-a,duplantier1982a-a}. One then
might expect that above the upper critical dimension ($d_u=3$) some
type of self-consistent mean-field theory based upon a suitable
tricritical Landau-Ginzberg Hamiltonian
\cite{gennes1975a-a,lawrie1984a-a} would give a full description of
the transition, and hence conclude that in all dimensions $d> 3$ there
is a collapse transition from a swollen state to the globular state
with classical tricritical behaviour. In the correspondence
\cite{gennes1975a-a} between a Landau-type functional for a magnetic
system with a magnetisation $M$ of $n$ components (in the limit
$n\rightarrow 0$) and the polymer problem, the coefficient of the $M^4$
in the Landau functional maps to the second virial coefficient $B$ of the
polymer solution and so to the temperature $T$ of the polymer,
while the temperature of the magnet $\tau$ is related to the polymer
length $N$ through $N\sim(\tau -\tau_c)^{-1}$ where $\tau_c(B)$ is the
location of the phase transition line in the so-called symmetry plane
\cite{lawrie1984a-a} of the tricritical system. A finite polymer
length implies the corresponding magnet is at a high temperature $\tau
> \tau_c$. Hence, the finite-size scaling around the collapse
transition should be described by tricritical crossover.

The application of the mean-field theory of a tricritical point to
polymer collapse predicts that at the transition point the polymer
actually behaves as if it were a random walk ($\nu_\theta=1/2$), and
this point has been known as the $\theta$-point. Thermodynamically
($N=\infty$), one expects a weak transition with a jump in the
specific heat $\alpha=0$ (note that the thermodynamic polymer exponent
$\alpha$ is related to the shift exponent in tricritical theory
\cite{lawrie1984a-a}, itself not to be confused with the polymer
theory finite-$N$ scaling $\psi_p$ shift exponent). For finite polymer
length $N$ there is no sharp transition for an isolated polymer
(unless one examines a macroscopic number of such polymers) and so
this mean-field transition is rounded and shifted. In three dimensions
the application of various self-consistent mean-field like approaches
predicts that the second-order transition is rounded and shifted on the
same scale of $N^{-1/2}$, that is, the crossover exponent $\phi$ is
$1/2$, though strictly the power laws involved are modified via
renormalisation group arguments by confluent logarithms.  In four and
higher dimensions no confluent logarithms should be present and one
may expect pure mean-field behaviour with a crossover exponent of
$1/2$ \cite{gennes1975a-a} ($\phi_t$ is the analogous tricritical
exponent \cite{lawrie1984a-a}).

On the other hand, for $d>3$ Sokal \cite{sokal1994a-a} has pointed out
that the alternative method of analysing collapse which has been shown
to be equivalent to the field theoretic approach, namely the continuum
Edwards model, has difficulties: in fact, if one analyses the Edwards
model one finds the crossover exponent is given by $\phi_E=2-d/2$,
which for $d=4$ gives $\phi_E=0$! In passing we note here that the same
analysis predicts the shift of the $\theta$-point, defined say via the
universal ratio of the radius of gyration to the end-to-end distance
equalling its Gaussian value, should scale as $N^{-(d/2 -1)}$ so
$\psi_E=(d/2 -1)\neq\phi_E$. This difference between the shift and the
crossover exponent implies that strict crossover scaling has broken
down. Of course, the theoretical fact that the swollen phase should
also be Gaussian for $d>4$ does raise the suspicion that the analysis
of the Edwards model for polymer collapse may be subtle for $d>3$.

To resolve this question of the crossover scaling one may first be
tempted to try some simple scaling arguments as follows.  Around the
$\theta$-point ($T\rightarrow T_\theta$ and $N\rightarrow \infty$) a
crossover form \cite{gennes1979a-a} is generally predicted to be:
\begin{equation}
\label{scaling-form}
R_{N} \sim N^{1/2} \: {\cal F}((T_\theta-T) N^{\phi}).
\end{equation}
The mathematical understanding of these forms was recently re-examined
\cite{brak1995a-:a}. To match the behaviour of (\ref{collapse-size})
consider the behaviour of (\ref{scaling-form}) for $T< T_\theta$ and
fixed, so we need to find the behaviour of ${\cal F}(z)$ for
$z\rightarrow\infty$. Assuming a power law ${\cal F}(z) \sim z^b$, and
the mean-field density exponent $\beta=1$, one finds that $b=1/d$ and
more importantly that $\phi=(d/2 -1)$. So we notice that the crossover
exponent found from this simple scaling argument is the same as
predicted from the Edwards model for the shift exponent. Hence we
might conclude that for $d>3$ the crossover scaling form might still
be applicable but with shift and crossover exponents that obey $\phi=
\psi_p= (d/2 -1)$. While this implies a subtlety in the Edwards model
analysis and also one in the mapping of the tricritical theory to the
polymer problem, it is perhaps not completely surprising.

\begin{figure}[h]
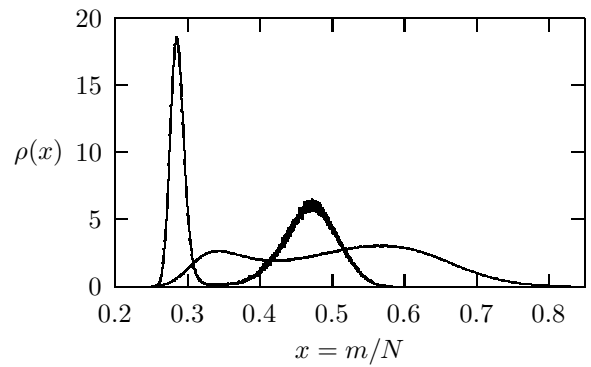

\begin{center}
\setlength{\unitlength}{0.240900pt}
\ifx\plotpoint\undefined\newsavebox{\plotpoint}\fi
\sbox{\plotpoint}{\rule[-0.200pt]{0.400pt}{0.400pt}}


\vspace{3mm}
\caption{\it Internal energy density distributions for lengths $N=2048$ and $16384$, each at
their respective transition temperatures. The more highly peaked distribution
is associated with length $16384$.}
\end{center}
\end{figure}

To consider such issues we have simulated interacting self-avoiding
walks, the canonical lattice model of polymer collapse, on the
four-dimensional hyper-cubic lattice using the PERM algorithm
\cite{grassberger1997a-a} over a wide range of temperatures with
surprising results. Because this algorithm is based upon kinetic
growth it works well around the collapse region as well as throughout
the swollen phase. As a consequence we are able to obtain reasonable
data up to length $N=16384$ (a more complete discussion of our
simulations is available in
\cite{prellberg1999=a-:a}). Now, firstly, our results suggest that
there is indeed a collapse transition in four dimensions at a finite
temperature. However, the character of that transition is particularly
intriguing!  In $d=4$ our scaling argument above predicts
$\phi=\psi_p=1$ while the mapping to the tricritical theory predicts
$\phi=1/2$: we find neither! In fact we find a rounded transition with
a divergent specific heat, and near the transition the distribution of
internal energy is clearly bimodal. This `double-peaked' distribution
becomes \emph{more} pronounced as the chain length is increased (see
figure 1). As we vary temperature through the transition region both
peaks are essentially stationary and one grows as the other decreases
in size: classic first-order behaviour. 
On the other hand we also were able to find a candidate $\theta$-point
(a critical state) where $R\sim N^{1/2}$ well above the transition
region. We then considered the shift of the rounded first-order like
transition to the $\theta$-point: the best scaling produced a shift
exponent of about $1/3$ (see figure 2).

\begin{figure}[h]
\begin{center}
\setlength{\unitlength}{0.240900pt}
\ifx\plotpoint\undefined\newsavebox{\plotpoint}\fi
\sbox{\plotpoint}{\rule[-0.200pt]{0.400pt}{0.400pt}}
\begin{picture}(900,585)(0,0)
\font\gnuplot=cmr10 at 10pt
\gnuplot
\sbox{\plotpoint}{\rule[-0.200pt]{0.400pt}{0.400pt}}
\put(140.0,123.0){\rule[-0.200pt]{4.818pt}{0.400pt}}
\put(120,123){\makebox(0,0)[r]{0.80}}
\put(859.0,123.0){\rule[-0.200pt]{4.818pt}{0.400pt}}
\put(140.0,229.0){\rule[-0.200pt]{4.818pt}{0.400pt}}
\put(120,229){\makebox(0,0)[r]{0.90}}
\put(859.0,229.0){\rule[-0.200pt]{4.818pt}{0.400pt}}
\put(140.0,334.0){\rule[-0.200pt]{4.818pt}{0.400pt}}
\put(120,334){\makebox(0,0)[r]{1.00}}
\put(859.0,334.0){\rule[-0.200pt]{4.818pt}{0.400pt}}
\put(140.0,440.0){\rule[-0.200pt]{4.818pt}{0.400pt}}
\put(120,440){\makebox(0,0)[r]{1.10}}
\put(859.0,440.0){\rule[-0.200pt]{4.818pt}{0.400pt}}
\put(140.0,545.0){\rule[-0.200pt]{4.818pt}{0.400pt}}
\put(120,545){\makebox(0,0)[r]{1.20}}
\put(859.0,545.0){\rule[-0.200pt]{4.818pt}{0.400pt}}
\put(140.0,123.0){\rule[-0.200pt]{0.400pt}{4.818pt}}
\put(140,82){\makebox(0,0){0.000}}
\put(140.0,525.0){\rule[-0.200pt]{0.400pt}{4.818pt}}
\put(288.0,123.0){\rule[-0.200pt]{0.400pt}{4.818pt}}
\put(288,82){\makebox(0,0){0.002}}
\put(288.0,525.0){\rule[-0.200pt]{0.400pt}{4.818pt}}
\put(436.0,123.0){\rule[-0.200pt]{0.400pt}{4.818pt}}
\put(436,82){\makebox(0,0){0.004}}
\put(436.0,525.0){\rule[-0.200pt]{0.400pt}{4.818pt}}
\put(583.0,123.0){\rule[-0.200pt]{0.400pt}{4.818pt}}
\put(583,82){\makebox(0,0){0.006}}
\put(583.0,525.0){\rule[-0.200pt]{0.400pt}{4.818pt}}
\put(731.0,123.0){\rule[-0.200pt]{0.400pt}{4.818pt}}
\put(731,82){\makebox(0,0){0.008}}
\put(731.0,525.0){\rule[-0.200pt]{0.400pt}{4.818pt}}
\put(879.0,123.0){\rule[-0.200pt]{0.400pt}{4.818pt}}
\put(879,82){\makebox(0,0){0.010}}
\put(879.0,525.0){\rule[-0.200pt]{0.400pt}{4.818pt}}
\put(140.0,123.0){\rule[-0.200pt]{178.025pt}{0.400pt}}
\put(879.0,123.0){\rule[-0.200pt]{0.400pt}{101.660pt}}
\put(140.0,545.0){\rule[-0.200pt]{178.025pt}{0.400pt}}
\put(509,21){\makebox(0,0){$N^{-2/3}\rule{7mm}{0pt}$}}
\put(436,281){\makebox(0,0)[l]{$N^{1/3}(\omega_{c,N}-\omega_\theta)\rule{20mm}{0pt}$}}
\put(140.0,123.0){\rule[-0.200pt]{0.400pt}{101.660pt}}
\put(867,510){\usebox{\plotpoint}}
\multiput(861.51,508.92)(-1.532,-0.499){173}{\rule{1.323pt}{0.120pt}}
\multiput(864.25,509.17)(-266.255,-88.000){2}{\rule{0.661pt}{0.400pt}}
\multiput(592.83,420.92)(-1.437,-0.499){115}{\rule{1.246pt}{0.120pt}}
\multiput(595.41,421.17)(-166.414,-59.000){2}{\rule{0.623pt}{0.400pt}}
\multiput(423.91,361.92)(-1.415,-0.498){73}{\rule{1.226pt}{0.120pt}}
\multiput(426.45,362.17)(-104.455,-38.000){2}{\rule{0.613pt}{0.400pt}}
\multiput(317.88,323.92)(-1.123,-0.497){57}{\rule{0.993pt}{0.120pt}}
\multiput(319.94,324.17)(-64.938,-30.000){2}{\rule{0.497pt}{0.400pt}}
\put(867,510){\circle{18}}
\put(598,422){\circle{18}}
\put(429,363){\circle{18}}
\put(322,325){\circle{18}}
\put(255,295){\circle{18}}
\end{picture}

\vspace{3mm}
\caption{\it Scaling of the shift of the transition: we show the scaling combination
$N^{1/3}(\omega_{c,N}-\omega_\theta)$ versus $N^{-2/3}$, where
$\omega_{c,N}$ and $\omega_\theta$ are the Boltzmann weights
associated with the monomer-monomer interaction at $T_{c,N}$ and
$T_\theta$.}
\end{center}
\end{figure}
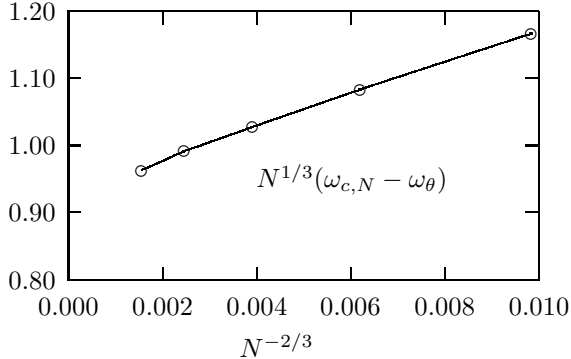

By studying the position when the universal ratio of the mean square
distance of a monomer to the end-point to the mean square end-to-end
distance takes on its Gaussian value we find that the $\theta$-point
is shifted much less and may scale as $1/N$. Hence there seem to be
\emph{two} shift exponents! While these results seem at variance with
standard tricritical ideas there is a mean-field type theory that
describes the first-order transition region well.  This framework was
explained some time ago in a paper by Khokhlov
\cite{khokhlov1981a-a}, who applied the mean-field approach of
Lifshitz, Grosberg and Khokhlov 
\cite{lifshitz1968a-a,lifshitz1976a-a,lifshitz1978a-a} to arbitrary
dimensions. Here we argue that the conclusions of these works may be
valid for $d\geq4$.  The theory is based on a phenomenological free
energy in which the competition between a bulk free energy of a dense
globule and its surface tension drive the transition. The consequences
of this surface free energy were largely ignored in the polymer
literature until recently, when its effect on the scaling form of the
finite-size partition function was proposed and confirmed
\cite{owczarek1993c-:a,owczarek1993d-:a,grassberger1995a-a,nidras1996a-a}.

Lifshitz theory \cite{lifshitz1968a-a} is based on several
phenomenological mean-field assumptions. Firstly, there exists a
$\theta$-state.  Secondly, for lower temperatures there exists a
globular state where the polymer behaves as a liquid drop. The results
of the theory are based on a phenomenological free energy of that
globular state relative to the free energy of the pure Gaussian state
of the $\theta$-point at $T_\theta$. The starting point of this
analysis is a bulk free energy with a quadratic dependence on the
distance to the $\theta$-point and so an exponent $\alpha=0$, implying
a second-order phase transition occurs in the thermodynamic
limit. From the theory one finds a rounded transition for finite $N$
occurring at temperature $T_{c,N} < T_\theta$ with
\emph{shift} exponent
\begin{equation}
\psi_p=1/(d-1)
\end{equation}
so that the collapse occurs at finite length at a temperature that
scales towards the $\theta$-point quite slowly and is below the
$\theta$-point.  This then concurs with our finding that $\psi_p
\approx 1/3$ in $d=4$. The width of this transition $\Delta T$ at
finite $N$ can be found and has \emph{crossover} exponent 
\begin{equation}
\phi=(d-2)/(d-1).
\end{equation}
All the exponents are derived from the assumptions of mean field
thermodynamic behaviour and of eq.\ (\ref{collapse-size}). Hence all
the exponents quoted here are related to each other (only one
independent exponent).  Our simulations show just such a narrow
crossover region with a crossover exponent that is certainly larger
than the shift exponent (the data is compatible with exponent $2/3$ in
$d=4$).

By considering the difference of the density $\rho_s$ of the swollen
state to the globular state $\rho_g$ relative to the density of the
swollen state at $T_{c,N}$, Khokhlov \cite{khokhlov1981a-a} concluded
from its divergence that `the coil-globule transition is first-order',
though we now interpret this to mean that the finite-size corrections
to the thermodynamic second-order transition are first-order like. We
point out that the terminology of Khokhlov was presumably that
explained in Section I.C.2 of
\cite{lifshitz1978a-a} but may be misleading to the modern
reader. However, both $\rho_g(T_{c,N})$ and $\rho_s(T_{c,N})$ tend to
zero as $N \rightarrow \infty $ and it is simply that
$\rho_g(T_{c,N})$ tends to zero asymptotically slower than
$\rho_s(T_{c,N})$ that makes the relative difference diverge. The
analysis can be used to deduce the scaling of $R_N$ at $T_{c,N}$ with
an exponent $\nu_c= 1/(d-1)$. Note that $\nu_\theta > \nu_c >
\nu_g$, so that this scaling is in-between the scaling fixed at the
$\theta$-point and at any temperature fixed in the collapsed
phase. Following the work of Lifshitz, Grosberg and Khokhlov
\cite{lifshitz1978a-a} one can also calculate the change in the
internal energy over the crossover width of the transition $\Delta T$
as the latent heat $\Delta U $ from the free energy expression. The
latent heat decays as $N$ increases with exponent $1/(1-d)$. The
corresponding height of the peak in the specific heat diverges with
exponent $(d-3)/(d-1)$. From our simulational data we were
unfortunately unable to extract reasonable estimates for $\nu_c$ or
the exponents of the latent and specific heats.

To interpret our results we can take the understanding of this
mean-field theory further. Let us consider the distribution of
internal energy, which we measured in our simulations, as a function
of temperature and length. For any temperature above the
$\theta$-point and well below $T_{c,N}$ one expects the distribution
of internal energy to look like a single peaked distribution centred
close to the thermodynamic limit value: a Gaussian distribution is
expected around the peak with variance $O(N^{-1/2})$. In fact this
picture should be valid for all temperatures outside a range of order
$O(N^{-(d-2)/(d-1)})$ centred on $T_{c,N}$.
When we enter this region we will expect to see a double peaked
distribution as in a first-order transition region. For any
temperature in this transition region there should be two peaks in the
internal energy distribution separated at the order of
$O(N^{-1/(d-1)})$ (the value of the gap being the latent heat).
Each of these peaks should be of Gaussian type with individual
variances again of the order $O(N^{-1/2})$. Hence as $N$ increases the
peaks will become more and more distinct and relatively sharper but
the peak positions will be getting closer together. Hence we refer to
this scenario as a \emph{pseudo}-first-order transition or, more
correctly, as first-order-like finite-size corrections to a
second-order phase transition. If there were a real first-order
transition then the distance between the peaks should converge to a
non-zero constant.  A comprehensive interpretation of our computer
simulations in $d=4$ \cite{prellberg1999=a-:a} is most consistent with
just such a scenario and leads us to conjecture that this theoretical
picture is indeed correct for the coil-globule transition for $d\geq
4$.

Let us return to the question of crossover scaling forms and
our finding of two shift exponents from our computer
simulations. While we cannot ascertain either with great accuracy let
us assume that we have a region around the $\theta$-point that is
approximated well by a form like (\ref{scaling-form}) with crossover
exponent $\phi_\theta=(d/2-1)$. Now, despite the fact that this does
not describe the collapse transition region, we notice that
substituting $t \sim N^{1/(d-1)}$, and using the asymptotics derived
from our matching argument above, leads to $R \sim N^{1/(d-1)}$, which
is precisely the correct scaling for the real transition region!
Hence we conjecture a phenomenological \emph{product} scaling form
(for $T< T_\theta$)
\begin{equation}
\label{product-scaling-form}
R_{N} \sim N^{1/2} \: {\cal F}((T_\theta-T) N^{\phi_\theta}){\cal
G}((T_{c,N}-T) N^{\phi})
\end{equation}
with $T_{c,N} \sim T_\theta - a N^{-\psi_p}$ and where ${\cal G}(y)
\sim 1$ for $y\rightarrow \pm\infty$.  This form will then correctly
describe both the region around the $\theta$-point and the rounded
transition around $T_{c,N}$ and will match with the behaviour of the
collapsed phase for fixed $T< T_\theta$. Such a form is not dependent
on the finding of pseudo-first-order behaviour and may be useful for
analysing data whenever two shift exponents are found. Two shift
exponents may appear in systems that are described by mean-field
theories.

Finally let us re-examine the crossover exponent derived from the
Edwards model. This is essentially only valid for $T>T_\theta$. For
$d\geq 5$ the swollen phase is Gaussian itself and the polymer
temperature is an irrelevant variable ($\phi_E <0$), and in $d=4$ it
is marginal, so one might suspect that the scaling theory for
$T>T_\theta$ is very different from that for $T<T_\theta$. Hence while
the scaling form (\ref{scaling-form}) from the Edwards model with
$\phi=\phi_E$ correctly predicts that $R \sim N^{1/2} f(T)$ for $d\geq
5$ it does not describe the collapse.

In conclusion, Monte Carlo simulations of lattice polymers in four
dimensions show that for finite length the rounded coil-globule
transition appears first order but we argue that the tricritical
predictions may well reappear in the infinite length limit as our
results fit best the predictions of Lifshitz-Grosberg-Khokhlov (LGK)
theory applied to high dimensions. We suggest that the crossover
scaling forms are more complicated than at low dimensions and suggest
a generalisation that may be heuristically useful. The Edwards model
is valid above the $\theta$-point while the coil-globule transition is
described by LGK theory.

\paragraph*{Acknowledgements} 
Financial support from the Australian Research Council is gratefully
acknowledged by ALO. This work was partially supported by EPSRC Grant No.\
GR/K79307 and a visiting scholars award from the University of
Melbourne's Collaborative Research Grants scheme. We thank
A. J. Guttmann for many useful comments on the manuscript.


\begin{thebibliography}{10}

\bibitem{lawrie1984a-a}
I.~D. Lawrie and S. Sarbach,  in {\em Phase Transitions and Critical
  Phenomena}, edited by C. Domb and J.~L. Lebowitz (Academic, London, 1984),
  Vol.~9.

\bibitem{gennes1975a-a}
P.-G. de~Gennes, J. Physique Lett. {\bf 36},  L55  (1975).

\bibitem{gennes1978a-a}
P.-G. de~Gennes, J. Physique Lett. {\bf 39},  L299  (1978).

\bibitem{gennes1979a-a}
P.-G. de~Gennes, {\em Scaling Concepts in Polymer Physics} (Cornell University
  Press, Ithaca, 1979).

\bibitem{duplantier1982a-a}
B. Duplantier, J. Physique {\bf 43},  991  (1982).

\bibitem{sokal1994a-a}
A.~D. Sokal, Europhys. Lett. {\bf 27},  661  (1994).

\bibitem{brak1995a-:a}
R. Brak and A.~L. Owczarek, J. Phys. A. {\bf 28},  4709  (1995).

\bibitem{grassberger1997a-a}
P. Grassberger, Phys. Rev. E {\bf 56},  3682  (1997).

\bibitem{prellberg1999=a-:a}
T. Prellberg and A.~L. Owczarek, in preparation (unpublished).

\bibitem{khokhlov1981a-a}
A.~R. Khokhlov, Physica A. {\bf 105},  357  (1981).

\bibitem{lifshitz1968a-a}
I.~M. Lifshitz, Zh. Eksp. Teor. Fiz. {\bf 55},  2408  (1968), [Sov. Phys. -
  JETP, \textbf{28}, 545, 1969].

\bibitem{lifshitz1976a-a}
I.~M. Lifshitz, A.~Y. Grosberg, and A.~R. Khokhlov, Zh. Eksp. Teor. Fiz. {\bf
  71},  1634  (1976), [Sov. Phys. - JETP, \textbf{44}, 1976].

\bibitem{lifshitz1978a-a}
I.~M. Lifshitz, A.~Y. Grosberg, and A.~R. Khokhlov, Rev. Mod. Phys. {\bf 50},
  683  (1978).

\bibitem{owczarek1993c-:a}
A.~L. Owczarek, T. Prellberg, and R. Brak, Phys. Rev. Lett. {\bf 70},  951
  (1993).

\bibitem{owczarek1993d-:a}
A.~L. Owczarek, J. Phys. A. {\bf 26},  L647  (1993).

\bibitem{grassberger1995a-a}
P. Grassberger and R. Hegger, J. Chem. Phys. {\bf 102},  6881  (1995).

\bibitem{nidras1996a-a}
P.~P. Nidras, J. Phys. A. {\bf 29},  7929  (1996).

\end{thebibliography}
\end{document}